\documentclass{elsarticle}
\usepackage{amsmath}
\usepackage{amssymb}
\usepackage{graphicx}
\usepackage{epsfig}
\usepackage{subfigure}

\def\Real{{\rm I\mathchoice{\kern-0.70mm}{\kern-0.70mm}{\kern-0.65mm}%
  {\kern-0.50mm}R}}

\font \bolditalics = cmmib10
\def\bx#1{\leavevmode\thinspace\hbox{\vrule\vtop{\vbox{\hrule\kern1pt
        \hbox{\vphantom{\tt/}\thinspace{\bf#1}\thinspace}}
      \kern1pt\hrule}\vrule}\thinspace}

\def \vc #1{{\textfont1=\bolditalics \hbox{$\bf#1$}}}

\def\kg{{\bf k}}

\def\thetag{{\vc \theta}}

\def\be{\begin{equation}}
\def\ee{\end{equation}}
\def\ba{\begin{eqnarray}}
\def\ea{\end{eqnarray}}

\def\mnras{{\em MNRAS}}
\def\apj{{\em ApJ}}                 
\def\apjl{{\em ApJ}}                
\def\apjs{{\em ApJS}}               
\def\ao{{\em Appl.~Opt.}}           
\def\aap{{\em A\&A}}                
\def\mnras{{\em MNRAS}}             

\begin{document}
\begin{frontmatter}
\title{Breaking the Degeneracy: Optimal Use of Three-point Weak Lensing Statistics}
\author{ Sanaz Vafaei$^1$, Tingting Lu$^2$, Ludovic van Waerbeke$^1$, Elisabetta Semboloni$^3$, Catherine Heymans$^{4,}$, Ue-Li Pen$^5$}
\address{$^1$Department of Physics and Astronomy, University of British Columbia, 6224 Agricultural Road, Vancouver, BC V6T 1Z1, Canada}
\address{$^2$Department of Astronomy and Astrophysics, University of Toronto, Canada M5S 3H4}
\address{$^3$Argelander-Institut f$\ddot{u}$r Astronomie, Auf dem H$\ddot{u}$gel 71, Bonn D-53121, Germany}
\address{$^4$Scottish Universities Physics Alliance (SUPA), Institute for Astronomy, University of Edinburgh, Royal Observatory, Blackford Hill, Edinburgh EH9 3HJ}
\address{$^5$Canadian Institute for Theoretical Astrophysics, University of Toronto, 60 St George Street, Toronto, Canada ON M5S 3H8}
\ead{svafaei@phas.ubc.ca}
\begin{abstract}
We study the optimal use of three-point statistics in the analysis
of weak lensing by large-scale structure. The three-point statistics
statistics have long been advocated as a powerful tool to break
measured degeneracies between cosmological parameters.  Using
ray-tracing simulations, incorporating important survey features
such as a realistic depth-dependent redshift distribution, we find
that a joint two- and three-point correlation function analysis is a
much stronger probe of cosmology than the skewness statistic.  We
compare different observing strategies, showing that for a limited
survey time there is an optimal depth for the measurement of
three-point statistics, which balances statistical noise and cosmic
variance against signal amplitude.   We find that the chosen CFHTLS
observing strategy was optimal and forecast that a joint two- and
three-point analysis of the completed CFHTLS-Wide will constrain the
amplitude of the matter power spectrum $\sigma_8$ to 10\% and the
matter density parameter $\Omega_m$ to 17\%, a factor of 2.5
improvement on the two-point analysis alone. Our error analysis
includes all non-Gaussian terms, finding that the coupling between cosmic 
variance and shot noise is a non-negligible contribution which should 
be included in any future analytical error calculations.

\end{abstract}
\end{frontmatter}

\section{\label{sec:Introduction}Introduction}
Weak gravitational lensing by large scale structure is a unique tool
to probe the matter distribution of the Universe regardless of its
dynamical state. When combined with redshift information weak lensing can be used as a probe for dark energy evolution as the expansion of the Universe affects the mass clustering at different redshifts. Dark energy constraints from weak lensing rely on accurate measurements of dark matter power spectrum amplitude.
The two-point cosmic shear statistics offer a
powerful technique to measure the matter
normalization parameter $\sigma_8$ and the mass density parameter
$\Omega_m$ combined (see for example the recent work on the CFHTLS
by \cite{Benjamin07} and \cite{Fu08}). 
One of the important goals for better determination of the cosmological parameters is to improve the individual measurement of $\sigma_8$ and $\Omega_m$.
Better estimation of $\sigma_8$ and $\Omega_m$ allows for the alleviation of
the residual parameter degeneracies \cite{Komatsu09}. A noticeable example is the neutrino mass \cite{Tereno08}.

Three-point shear statistics, in particular the skewness, are in principle
powerful estimators to break the degeneracy between $\sigma_8$ and
$\Omega_m$ (\cite{Bernardeau97} and \cite{waerbeke99}). Some
detections have been reported from the VIRMOS survey
(\cite{Bernardeau03} and \cite{Pen03}) and from the CTIO survey
\cite{Jarvis05}.
Unfortunately, the signal-to-noise ratio remains low and there are
still no reliable forecasts of three-point statistics which take
into the account realistic galaxy number counts and shape noise as
well as non-Gaussian contributions in the cosmic variance. Therefore,
the interpretation of the measurement is currently not
well-established. In \cite{waerbeke01} the authors concluded that
the three-point statistics of lensing signal is greatly enhanced at small angular
scales because of the non-linear gravitational clustering, but they
did not provide an estimate of the signal-to-noise ratio for
different survey depths. In \cite{waerbeke99} it was shown that the
skewness of the convergence can be measured from mass maps
reconstructed from the shear measured on individual galaxies,
however, a realistic population of source galaxies was not
considered, and the simulations were limited to second order
perturbation theory. \cite{Kilbinger05} showed that one can learn
additional information by combining the two- and three-point
statistics, but again neither a realistic source galaxy distribution
nor different survey strategies were considered. \cite{Takada04} also showed that combining the power spectrum and bispectrum tomography information enhances the accuracy of cosmological parameter estimations.

In this paper we investigate the optimal use of three-point statistics in a weak lensing analysis of large scale structure, considering several new aspects that have been
neglected in previous works:
\noindent
\begin{itemize}
\item A realistic noise contribution using ray-tracing simulations calibrated on
existing surveys is included.
\item Realistic forecasting for the two- and three-point statistics for different survey strategies is provided.
\item For a fixed observing time, wide-shallow and narrow-deep strategies are
considered. 
The impact of the survey's depth on both the galaxy number density and the source redshift distribution is quantified. Surveys with different
 characteristics are  affected differently by  cosmic variance, with wider
surveys probing a much larger number of modes than narrow surveys. Here we
carefully investigate this aspect by comparing the performance of various
simulated surveys which use a realistic source distribution.
\item The source distribution has been derived using  galaxy counting  as a function of redshift as measured in real data for a  fixed limiting magnitude.
\item The full likelihood analysis with covariance matrices are computed from a large set of ray-tracing simulations. It is therefore an extension of previous works which used fisher matrices
to gauge the performance of weak lensing surveys (e.g. \cite{Amara07}).
\item Following \cite{Zhang03} a comparison of different smoothing filters is included. 
\item A range of most optimal smoothing scales are found by investigating the various contributions of noise and signal to the full covariance matrix.
\item The best survey strategy for detecting the skewness of the convergence $S_3$ as means of breaking the degeneracy between $\Omega_m$ and $\sigma_8$ is studied. The idea first emerged in \cite{Bernardeau97} and \cite{waerbeke99} but its feasibility never quantified.
\item The efficiency of combining the two- and three-point statistics is quantified.
\item Two- and three-point statistics forecasts for the completed CFHTLS survey and the KIDS survey are calculated.
\end{itemize}

The paper is organised as follows. In \S\ref{sec:theory}, we
summarize the background theory of the two- and three-point
statistics of the  convergence field, where notations and
definitions are also introduced. The details of the method are
described in \S\ref{sec:method}. Optimal survey strategies are shown
in \S \ref{sec:optimal}, and \S \ref{sec:CFHTLS} shows the
predictions of two- and three-point measurements of the simulated
complete CFHTLS-Wide survey area and depth. The
upcoming KIlo Degree Survey geometry is also discussed here as an example of the accuracy achievable
on the measurement of the two- and three-point statistics in the near future.
Finally in \S \ref{sec:conclusion} the conclusions of this
study are stated.

\section{\label{sec:theory}Theory Background}
Following  \cite{Miralda91} and
\cite{Kaiser92} we can write the convergence $\kappa$ at a given
sky position $\theta$ as :

\begin{eqnarray}
\kappa(\thetag)&=&\frac{3}{2}\frac{H_0^2}{c^2}\Omega_{m}\int_{0}^{\infty}\omega(z)\delta(\chi,\thetag)d\chi
\label{kappadef}
\end{eqnarray}
where $\chi$ is the angular comoving distance, $\Omega_{m}$ is the
mass density parameter at the present day, $\delta$ is the matter
density contrast and $\omega(z)$ for a given redshift $z$ is given
by:
\begin{eqnarray}
\omega(z)&=&(1+z)\chi(z)\int_{z}^{\infty}n(z_{s})\left[{1-\frac{\chi(z)}{\chi(z_{s})}}\right]dz_{s}
\label{omegadef}
\end{eqnarray}

$\omega(z)$ depends on the cosmological
parameters and the galaxy source distribution function $n(z_s)$. The
convergence maps are obtained from ray-tracing simulations, as
described in Section \ref{sec:simulations}.

Note that this analysis  employs the convergence field $\kappa$,
which is proportional to the projected mass density. The convergence
can be obtained from the shear data $\gamma=(\gamma_1,\gamma_2)$
either by appropriate weighting with an aperture filter, or from
mass reconstruction with e.g. a top-hat or Gaussian filter.
Therefore, the conclusions of this study apply to the convergence
and the shear without distinction.
We are interested in the measurement of $\langle \kappa^2\rangle$,
$\langle \kappa^3\rangle$ and the skewness $S_3(\kappa)$  defined as:

\begin{equation}
S_3={\langle \kappa^3\rangle\over \langle \kappa^2\rangle^2}.
\label{eq:s3}
\end{equation}

 Skewness is essentially a measure of the clustering of the mass distribution as defined in \cite{Bernardeau97}. According to perturbation theory $S_3$
provides a measurement of $\Omega_m$ independent of the
normalization of power spectrum $\sigma_8$. For
this reason, the skewness of the convergence appears as a very
attractive probe of cosmology and a useful technique to break
degeneracies among other cosmological parameters.

\section{\label{sec:method}Analysis Method}

\subsection{\label{sec:simulations}Ray-tracing Simulations}

This analysis is based on  a set of simulated $\kappa$-maps, with 60
lines of sight each containing 40 redshift slices from $z=0.020$ to
$z=3.131$. These are generated from 22 independent N-body
simulations by randomization. As a result, the different lines of
sight are not totally independent on large scales. However they can
still be considered  approximately independent on scales smaller
than 1 degree.

The Multiple Lens-Plane ray-tracing approximation method was used to
generate the lensing convergence map: the dark matter distribution
in the universe is approximated by a series of mass sheets. The
N-body simulations are on a grid of $1728^3$ points with $856^3$
particles, and the box size is $120$$~$$h^{-1}$Mpc. The mass density
in the simulation box is projected to the mid-plane at a series of
characteristic redshifts. The output redshifts are picked so that
the consecutive time slices can represent the continuous evolution
of the large scale structure. The three orthogonal axes of the box
are $x$, $y$, and $z$. For every output redshift, we make three
projection sheets, parallel to the $xy$, $yz$ and $xz$ planes. We
choose one projection sheet out of the three of one N-body
simulation in a random order, as well as randomly shifting the sheet
transverse to the projection direction. This technique is employed
to avoid creating periodicity in the projection. Rays are shot
through these mass sheets. We calculate $\kappa$ on every sheet and
project them along lines of sight after the random shift and
rotation.

\begin{figure}
\begin{center}
\includegraphics[scale=0.6]{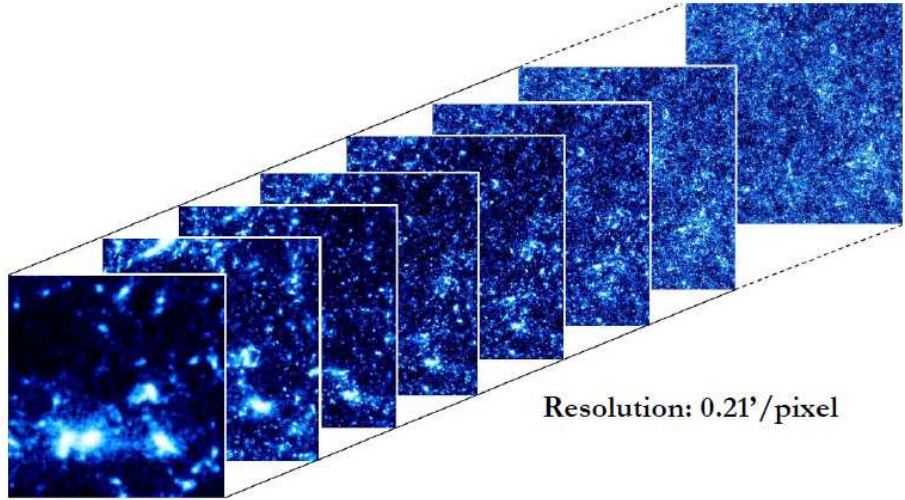}
\end{center}
\caption{A schematic of the simulated convergence maps at different redshift
slices.The maps are on 1024$^2$ with 0.21 arcmin per pixel. The
redshift ranges from z=0.020 to z=131. } \label{fig:blueslices}
\end{figure}

The maps are on 1024$\times$1024 grid with spacing of 0.21 arcmin.
Thus the total area is about 12.84 deg$^2$ for each line of sight.
We assume the following cosmology: 
$\Lambda_{\rm CDM}$ with
$\Omega_{m}=0.24$, $\Omega_{\Lambda}=0.76$ and $\sigma_{8}=0.74$ \cite{Spergel07}.
Figure \ref{fig:blueslices} shows a schematic of the different
redshift slices which were combined for each line of sight.

The N-body simulations are generated by the {\small CUBEPM} code,
which is the successor of {\small PMFAST} \cite{Merz05}. {\small
CUBEPM} is MPI parallelized particle-mesh (PM) code, and has
particle-particle force implement at sub-grid scales. It is further
parallelized by shared-memory OpenMP on each node. The simulation
volume (which is also called simulation box) is cubically
decomposed to $n^3$ sub-volumes, and the calculation of each
sub-volume is performed on one node of the cluster. The total number of
nodes used in simulation is $n^3$ with $n=3$ here. The code can be run on up to 1000
nodes. The simulations are run on the Sunnyvale cluster of CITA.

\begin{figure}
\begin{center}
\includegraphics[scale=0.8]{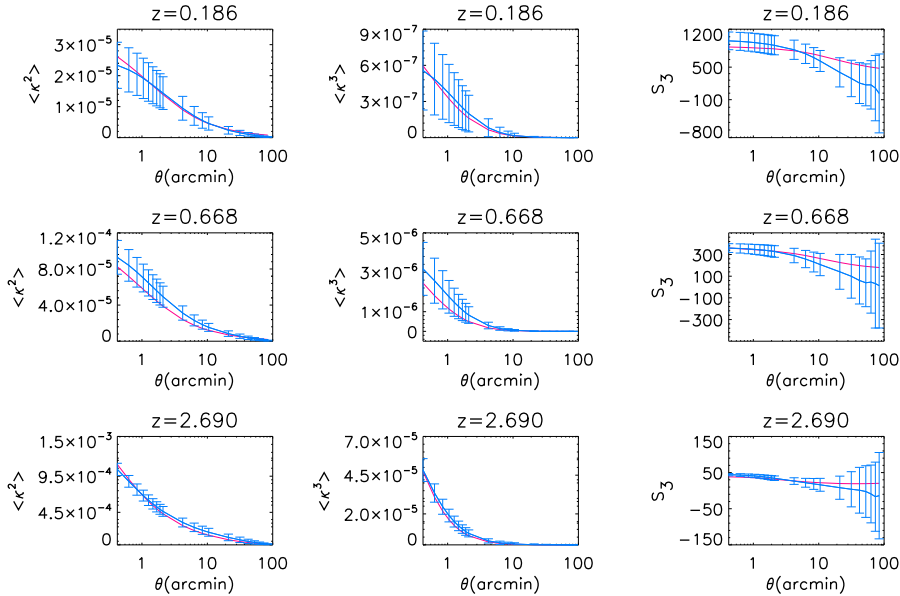}
\end{center}
\caption{The agreement between the measurements and the theoretical
predictions for three individual redshift slices. The low redshift
slice is at 0.186, the medium slice at 0.668 and the high redshift
slice is at $z$=2.690. The blue lines show the measurements on the
simulated 12.84 deg$^2$ data, and the pink lines show the theoretical
prediction for the same cosmological model at the same redshift. The
measurements in each panel are made from data smoothed with a top-hat
filter. The errorbars represent the cosmic variance over 60 lines of
sight. } \label{fig:oneslice}
\end{figure}

 For each of the redshift slices the average $\langle
\kappa^2 \rangle$, $\langle \kappa^3 \rangle$ and $S_3$ are measured and  the
signal is compared with a theoretical model. The
two-point cosmological predictions are based on the Peacock and
Dodds \cite{PeacockDodds86} non-linear fit, whereas the three-point
shear statistics predictions use the bispectrum non-linear fit
derived in \cite{Scoccimarro98} and implemented for lensing studies
in \cite{waerbeke01}. The excellent  agreement between the measured
 and the predicted signal can be seen in figure
 \ref{fig:oneslice} where the results for low, intermediate and high redshift slices are shown.

\subsection{\label{sec:galaxynumber}Galaxy Number Density and Redshift Distribution}
In this paper we compare different survey strategies with varying source redshift distribution that is dependent on the survey depth. We calibrate the redshift distribution from existing optical surveys with photometric redshift information and populate the ray-tracing slices accordingly.
The focus here is on ground based
surveys, but the result can be straightforwardly
extended to space data with an appropriate scaling of the shot noise
(which directly depends on the galaxy shape noise and number
density).

The galaxy number density and redshift distribution as a function of
limiting magnitude are estimated from the CFHTLS-Deep survey
catalogue in i-band \cite{Ilbert06}. To model the galaxy redshift distribution
$n(z)$ for surveys of different magnitude limit $\rm m_{lim}$, the method described in \cite{Heymans06} and \cite{waerbeke06} was employed, modeling $n(z,{\rm m_{lim}})$ as:

\begin{equation}
n(z,{\rm m_{lim}})= \frac{\beta}{z_0\Gamma\left(
\frac{1+\alpha}{\beta}\right)} \left( \frac{z}{z_0({\rm
m_{lim}})}\right)^{\alpha} \exp \left[ - \left( \frac{z}{z_0({\rm
m_{lim}})} \right)^{\beta}\right] \label{eq:nofz}
\end{equation}

\begin{figure}
\begin{center}
\includegraphics[scale=0.6]{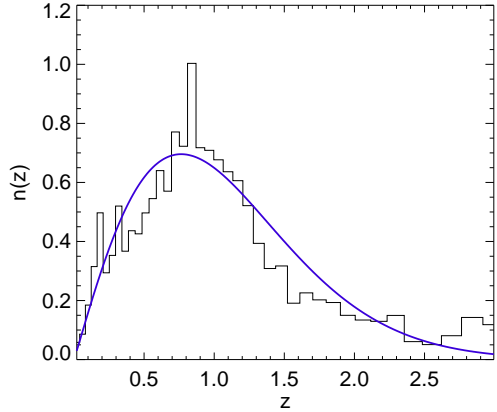}
\end{center}
\caption{The fit to the normalized galaxy number density from the
CFHTLS-Deep survey catalogue \cite{Ilbert06}. The black line shows
the histogram of the galaxy number density and the dark blue line is
the fitted curve. The limiting magnitude $\rm m_{lim}=$24.5 with the
fitting formula given by equation (\ref{eq:nofz}). Here  $\alpha$,
$\beta$ and $z_0$ are 0.96, 1.70 and 1.07 respectively. }
\label{fig:nofz}
\end{figure}

The best parametric fit to equation (\ref{eq:nofz}) for limiting
magnitude $i=$24.5 corresponds to  $\alpha$=0.96, $\beta$=1.70 and
$z_0$=1.07. Figure \ref{fig:nofz} shows the histogram of the
normalized galaxy redshift distribution from the CFHTLS-Deep survey
catalogue \cite{Ilbert06} at $\rm m_{lim}=$24.5 and the best fit
$n(z)$ from equation (\ref{eq:nofz}). Table \ref{tab:alfabetaz0}
summarizes the values of $\alpha$, $\beta$, $z_0$ and the median
redshift $z_{\rm{med}}$ for the other magnitude cuts used in this
paper.
Equation (\ref{eq:nofz}) yields a realistic source redshift distribution
for a given survey's depth \cite{waerbeke06}. Thus, the theoretical predictions built based on  the appropriate form of equation (\ref{eq:nofz}) match the $\kappa$-maps weighted by  the galaxy number density derived from the CFHTLS-deep catalogues.

\begin{table}
\centering
\begin{tabular}{l l l l l}
\hline
$\rm m_{lim}$&$\alpha$&$\beta$&$z_0$&$z_{\rm{med}}$\\
\hline
22.5&0.76&6.85&1.05&0.68\\
23.0&0.71&5.30&1.14&0.72\\
23.5&0.81&3.15&1.19&0.80\\
24.0&0.80&2.72&1.26&0.84\\
24.5&0.96&1.70&1.07&0.91\\
25.0&0.85&1.90&1.26&0.96\\
25.5&1.46&1.30&0.75&1.02\\
26.0&1.71&1.27&0.68&1.04\\
\hline
\end{tabular}
\caption{The best fit values of $\alpha$, $\beta$ and $z_0$
corresponding to equation (\ref{eq:nofz}) for several i-band limiting
magnitudes. These parameters were used to generate theoretical models
for Section  \ref{sec:optimal} to determine the best survey
strategy. The last column contains the median redshift $z_{\rm{med}}$ for
each magnitude cut.} \label{tab:alfabetaz0}
\end{table}

\subsection{\label{sec:noise} Statistical Noise}

The source of shot noise in weak lensing studies depends on the
intrinsic ellipticity characterized by the r.m.s. $\sigma_\epsilon$
and by the number density of galaxies $n_g$. It was shown in
\cite{waerbeke00} that the noise in a smoothed convergence map can
simply be derived from the intrinsic ellipticity noise and the
galaxy number density. In particular, it was shown that the noise in
a pixelated smoothed $\kappa$ map is simply given by a smoothed
uncorrelated Gaussian noise with r.m.s. $\sigma_\epsilon$. If $n_g$ denotes
the number density of galaxies and $W(\thetag)$ the 2-dimensional
smoothing function, then the correlation function of the convergence
noise is:

\begin{equation}
\langle
\kappa_n(\thetag)\kappa_n(\thetag')\rangle={\sigma_\epsilon^2\over
2}{ 1\over \Theta^2 n_g} \int d\kg e^{i\kg\cdot
\left(\thetag-\thetag'\right)}\left|\tilde W(\kg)\right|^2
\end{equation}
where $\kappa_n(\thetag)$ is the convergence noise map and $\tilde
W(\kg)$ is the Fourier transform of the smoothing window
$W(\thetag)$. $\Theta$ is the pixel size, so $\Theta^2 n_g$ is the
average number of galaxies per pixel.

The galaxy ellipticity r.m.s. measured on CFHTLS-deep data is
$\sigma_{\epsilon}^2=(\sigma_{\epsilon_1}^2+\sigma_{\epsilon_2}^2)=0.44$.
For the purpose of this paper we will assume that $\sigma_\epsilon$ is constant as a function of
redshift and galaxy type. The convergence noise variance per pixel
(before smoothing with $W$) is therefore given by:

\begin{equation}
\sigma_{\kappa}^2=\frac{\sigma_{\epsilon}^2}{2}\frac{1}{\Theta^{2}n_g}
\end{equation}
Note that the noise model considered here implicitly assumes sources
galaxies are distributed randomly in each redshift slice. By
construction, this choice ignores any potential effect caused by
source clustering, which is known to be a source of contamination
for three-point statistics  (\cite{Bernardeau98} and \cite{Forero07}).

\subsection{\label{sec:filters}Smoothing Filters}
\begin{figure}
\begin{center}
\includegraphics[scale=0.8]{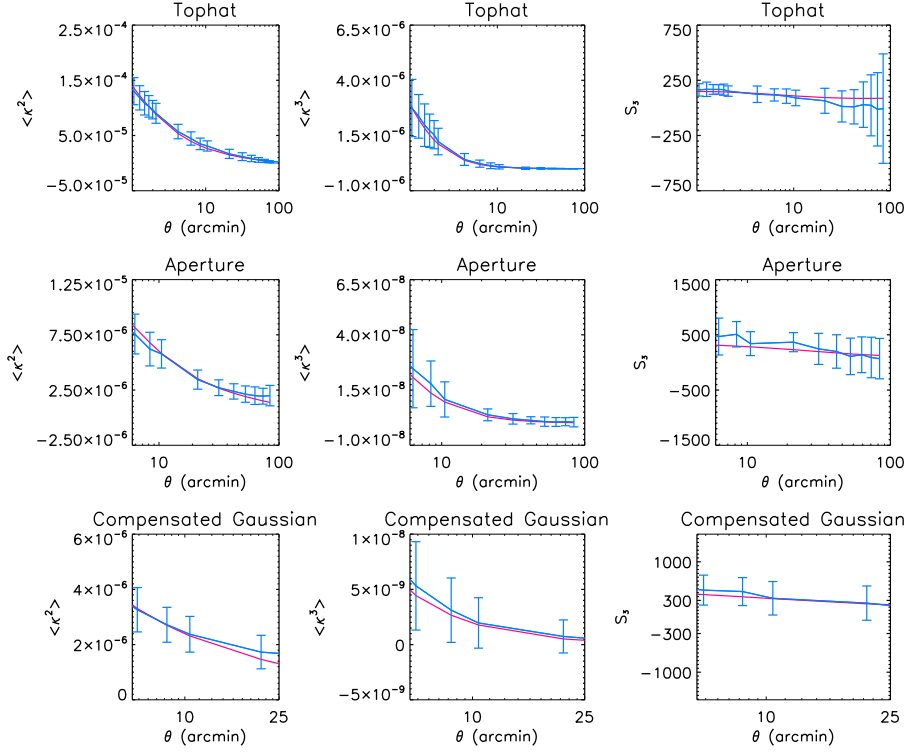}
\end{center}
\caption{The agreement between the measurements and the theoretical
predictions based on the fitted galaxy population. The blue lines
show the measurements on the simulated 12.84 deg$^2$ data, and the
pink lines show the theoretical prediction for the same cosmological
model and the full redshift distribution. The measurements in each
panel are performed on smoothed data, with top-hat, aperture mass and
Compensated Gaussian filters in order. The errorbars include both
cosmic variance and statistical noise resulting from $n_g$= 22 galaxies per arcmin$^2$. } \label{fig:match}
\end{figure}

 Convergence statistics can be  measured from
smoothed $\kappa$-maps (which can be obtained from smoothed shear maps from
the data). Various statistics can be build by using different smoothing filters. Following the widely accepted choice the top-hat, and two types of the
compensated (the total area under the filter window is equal to
zero) filters were considered. The two compensated
filters used were the ones introduced in \cite{Schneider98} (hereby
referred to as the aperture filter) and in \cite{waerbeke98} which
 hereby referred to as the Compensated Gaussian (CG). They are
defined as:
\begin{eqnarray}
W_{\rm ap}(\theta)&=&\frac{9}{\pi}\left( \frac{1}{\theta_f}\right
)^2 \left [1-\left (\frac{\theta}{\theta_f}\right )^2\right ]\left
[\frac{1}{3}-\left (\frac{\theta}{\theta_f}\right)^2\right] \text{
if } \theta < \theta_f  \text{ zero otherwise}\cr W_{\rm
CG}(\theta)&=&\frac{1}{2\pi{\theta_f}^2}\left[1-\left
(\frac{\theta^2}{2{\theta_f}^2}\right)\right]\exp\left(-\frac{\theta^2}{2{\theta_f}^2}\right)
\label{eq:filters}
\end{eqnarray}
where  $\theta_f$ is the characteristic size of the filter.
Figure \ref{fig:match} shows the excellent agreement between the ray-tracing
simulation and the predictions for different smoothing filters. The measurements are based on a realistic redshift distribution corresponding to a
ground based survey with limiting magnitude $m_{\rm lim}=24.5$ with $n_g$=22 galaxies per arcmin$^2$. The error bars reflect the statistical noise and cosmic variance for a $12.84$ deg$^2$ survey. From equation (\ref{eq:filters})  the smoothing scale for the two filters are related as $\rm{\theta}_{\rm {CG}}=\theta_{ap}/2\sqrt{2}$, therefore the maximum smoothing scale chosen for the Compensated Gaussian filter is 25 arcminutes compared to 84 arcminutes for the top-hat and aperture filters.

\subsection{\label{sec:CV}Estimate of the Covariance Matrix}

Cosmological parameter forecasting requires the estimate of the
covariance matrix. Semi analytical methods are available in the
literature (\cite{Schneider02} and \cite{Joachimi08}), but rely on
the assumption of Gaussian statistics. An extension to the
non-linear angular scales has been recently developed
(\cite{Semboloni07}, \cite{Eifler08} and \cite{Pielorz09}), however the three-point statistics and source redshift distribution and shape noise of realistic surveys were not considered.
 In this work the full covariance matrix {\bf C} was estimated directly
from the ray-tracing simulation as in \cite{Semboloni07} and
\cite{Eifler08}, by taking into account the realistic
characteristics of lensing surveys described in the previous
sections.  For each survey strategy, the total
covariance matrix was calculated as follows: for each noise-free $\kappa$ line of
sight, the redshift slices were combined and weighted according to
equation (\ref{kappadef}) with the corresponding redshift distribution and
galaxy number density. A noise map was then added following the
method described  in Section \ref{sec:noise}. Finally, the two- and
three-point statistics were measured over 20 smoothing scales.   The
covariance matrix of the statistic $x$  measured at  two smoothing
 scales  $\theta_i$ and
$\theta_j$ is defined as:
\begin{equation}
{\bf C}(\theta_i,\theta_j)\equiv \langle
(x(\theta_i)-\mu(\theta_i))(x(\theta_j)-\mu(\theta_j))\rangle
\label{eq:cov}
\end{equation}
where $x$  is here either  $\langle
\kappa^2 \rangle$, $\langle \kappa^3 \rangle$ or $S_3$ and $\mu$
is the average calculated from the entire simulation set.

It was shown by \cite{Hartlap07} that the inverse of the covariance
matrix estimated from a finite number of ray-tracing simulations is
biased. The authors derived a simple formula to correct for this
effect which relates the number of scales $p$ used in the two- (or
three-) point statistics and the number $n$ of lines-of-sights. The
covariance matrix simply has to be replaced by $\alpha^* {\bf C}$,
where $\alpha^*$ when the mean is determined from the data, is given
by:

\begin{equation}
\alpha^*=\frac{(n-1)}{(n-1)-p-1} \label{alphadef}
 \end{equation}
\cite{Hartlap07} showed that this correction is applicable only when
$n-2$ exceeds the number of scales $p$, otherwise the covariance
matrix ${\bf C}$ is not invertible. In this paper $n=60$ simulations
were used, and the statistics were measured over $p=20$ angular
scales for top-hat and aperture filter and $p=14$ for the Compensated
Gaussian filter. The values of $\alpha^*$ for these filters were then
$\rm 1.55$ and $\rm 1.28$ respectively.
For joint likelihood calculations the joint covariance matrices were rescaled by
$\alpha^*=3.28$ for top-hat and aperture filters and $\alpha^*=1.97$
for the Compensated Gaussian filter.

\begin{figure}[h!]
\center
\includegraphics[scale=0.5]{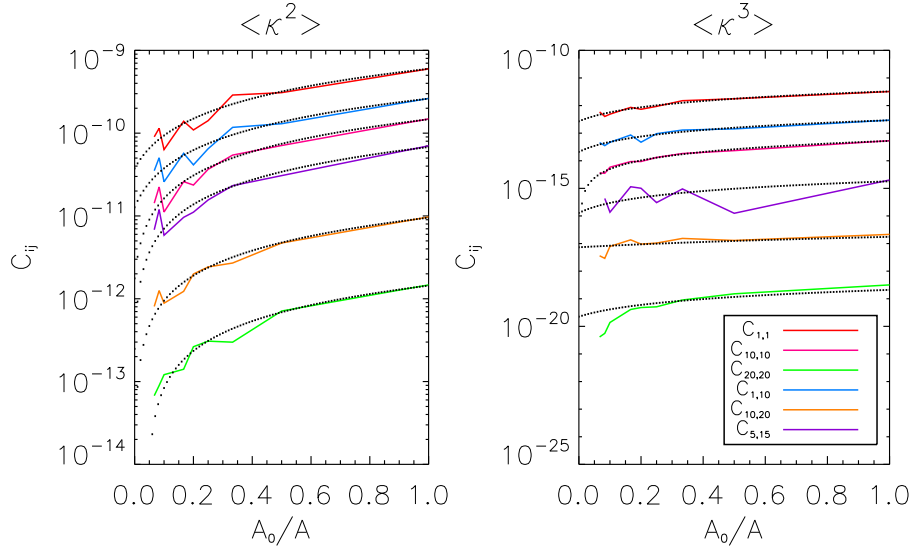}
\caption{The $\rm C_{ij}$ elements of the covariance matrix as a
function of the ratio of the original $\kappa$-map simulation area (12.84 deg$^2$) over the survey area. The solid colored lines are
the $\rm C_{ij}$ elements from the simulated maps and the black dotted
lines are the straight line fit to each of them. The left figure
shows the $\rm C_{ij}$ of $\langle \kappa^2 \rangle $ and the right one
is the same for  $\langle \kappa^3 \rangle $. Here the covariance
matrix contains only the cosmic variance. The scales are as
following: i=1 is 0.42', i=5 is 1.26', i=10 is 4.20', i=15 is 31.5'
and i=20 is 84.0'. It shows that the change in the covariance matrix
of the cosmic variance is inversely proportional to the survey area.
Hence  this result was used to rescale the covariance matrices
  in the likelihood calculation to the desired survey area.}
\label{fig:flatmatrix}
\end{figure}

Because of the limited area covered by the simulations it is not
possible to compute the covariance matrices for very large surveys.
Fortunately, the angular scales where the non-linear effects are
important (typically less than half a degree for the two- and three-
point statistics) are much smaller than the 12.84 deg$^2$
field-of-view of a simulation field. Those are also the scales where
the lensing signal is best measured. Therefore the covariance
matrices can be computed in the non-linear regime from the different
realizations, and simply rescaled according to survey size
for surveys exceeding the simulation box. This is an excellent
approximation for angular scales much smaller than the simulation
box, which was the case in our study since the largest scale used to
measure the statistics was $84$ arcminutes, which is much smaller
than the dimension of $3.5\times 3.5$ degrees of the simulations
box. Figure \ref{fig:flatmatrix} illustrates the scaling applied to
some elements of the two- and three-point statistic covariance
matrices.

\begin{figure}[h!]
\center
\includegraphics[scale=0.75]{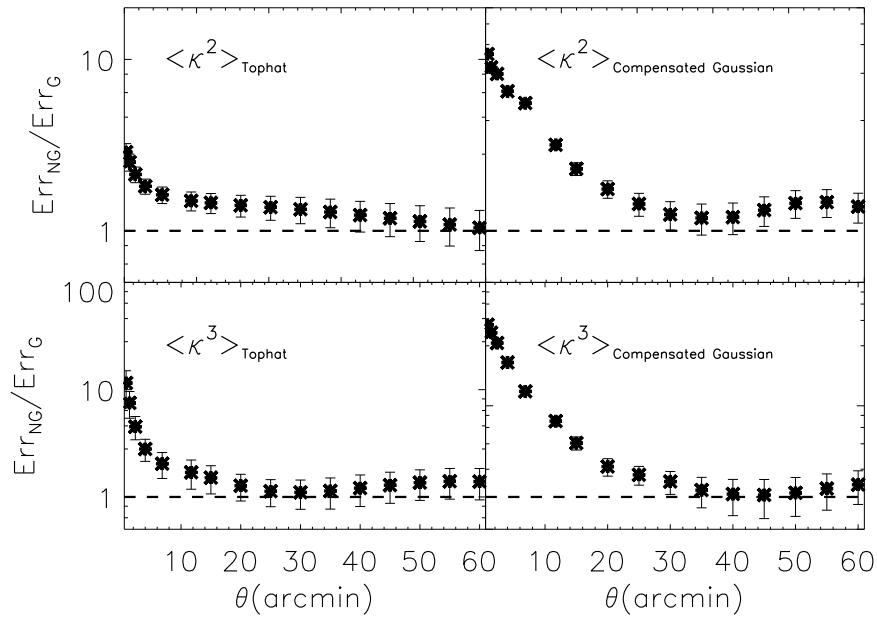}
\caption{The ratio of Gaussian to non-Gaussian error estimated for convergence $\kappa$ two- and three-point statistics. On large scales the non-Gaussian errors estimated from the ray-tracing simulations converge to the Gaussian limit. The results for the top-hat and Compensated Gaussian filters are shown in left and right column respectively.}
\label{fig:gaussian}
\end{figure}

In order to verify that the covariance matrix computed using ray-tracing simulations converges to the one computed in the Gaussian approximation for large angular scales, the following procedure was performed:  Gaussian realizations of the field $\kappa$ were generated and then the covariance matrix  was calculated in the same way as the ray-tracing simulations. \cite{Efstathiou85} and \cite{Salmon96} described a simple way to generate cosmological Gaussian fields by convolving white-noise with a filter whose transfer function is given by the square root of  the power spectrum. 
The power spectrum was directly computed from the sample of ray-tracing simulations, so that the resulting Gaussian fields have the same cosmology. Using this method, 60 lines of sight were generated and the covariance matrix of the Gaussian fields was computed as described by equation (\ref{alphadef}). Figure \ref{fig:gaussian} shows the ratio of the non-Gaussian to Gaussian errors (i.e. the square root of the diagonal elements of the covariance matrix)  for the two- and three-point statistics of the top-hat and Compensated Gaussian filters. 
It can be seen that for large scales the ratio converges to unity as expected. At small scales this ratio is larger than the unity due to the non-linear evolution of matter fluctuations. Moreover, for a given angular scale,  the ratio between non-Gaussian and Gaussian errors is larger  when one uses the Compensated Gaussian filter than when  the top-hat filter is used. The reason lies in the fact that for a given characteristic scale, the Compensated Gaussian filter peaks at smaller scales than the top-hat filter. The ratio between non-Gaussian and Gaussian covariance depends on the average redshift of the survey and for this test a distribution characterized by an average redshift z $\sim$ 1.4 was used. The ratio would have been much higher if a much shallower survey was chosen.

\section{\label{sec:optimal} Survey Design and Observing Strategy}

\subsection{\label{sec:optimalscale}Optimal smoothing scale}
The covariance matrix contains three terms (\cite{Schneider02}) :

\begin{equation}
{\bf C}={\bf C}_{\rm ss}+{\bf C}_{\rm ns}+{\bf C}_{\rm nn}
\label{covar}
\end{equation}

where ${\bf C}_{\rm ss}$ is the pure signal (i.e. noise free) cosmic
variance, ${\bf C}_{\rm nn}$ is the pure noise covariance and ${\bf
C}_{\rm ns}$ is the cross-correlation term.  The goal in this section is to  determine at which angular scale the measurement of the two- and three-point shear statistics  has a  better signal-to-noise ratio. For this purpose the covariance matrix was separated into the three terms introduced above and their amplitudes for different filters were explored.
Practically ${\bf
C}_{\rm ss}$ can be calculated from the noise free ray-tracing
realizations and ${\bf C}_{\rm nn}$ from noise-only convergence
maps. Among the three parts of the covariance matrix: ${\bf C}_{\rm
ss}$, ${\bf C}_{\rm nn}$, and ${\bf C}_{\rm ns}$, the mixed term
${\bf C}_{\rm ns}$ is the most computationally expensive to calculate. The reason is
that the noise contribution to the covariance matrix converges more
slowly than the cosmic variance contribution, and in practice, it is
necessary to estimate the noise from more than $60$ noise
realizations. For the two-point statistics there are analytical
formulae in \cite{Schneider02}, but there is currently no equivalent
for the three-point statistic and the skewness of the convergence.
In order to inspect the three different terms, the covariance matrix was calculated as follows:
for each noise realization, a ${\bf C}$ was calculated, which was relatively noisy because it was obtained from one noise pattern. Then the average of ${\bf C}$ was taken over ten noise realizations. The covariance matrix thus obtained was specialized for a given noise
statistical property, and the whole calculation was repeated
each time the observing conditions affecting the noise were changed.
 ${\bf C}_{\rm nn}$ was calculated separately over ten thousand realizations. The average ${\bf C}_{\rm ss}$ and ${\bf C}_{\rm nn}$ were used to determine the cross term ${\bf C}_{\rm ns}$.
Because of the averaging process one obtains a covariance matrix
which has a relatively small noise making ${\bf C}$ invertible. To
illustrate the contribution of each of these parts the
diagonal elements of the ${\bf C}_{\rm ss}$, ${\bf C}_{\rm nn}$,
${\bf C}_{\rm ns}$ and ${\bf C}$ were extracted as the noise term for each
smoothing scale.

\begin{figure}
\includegraphics[scale=0.6]{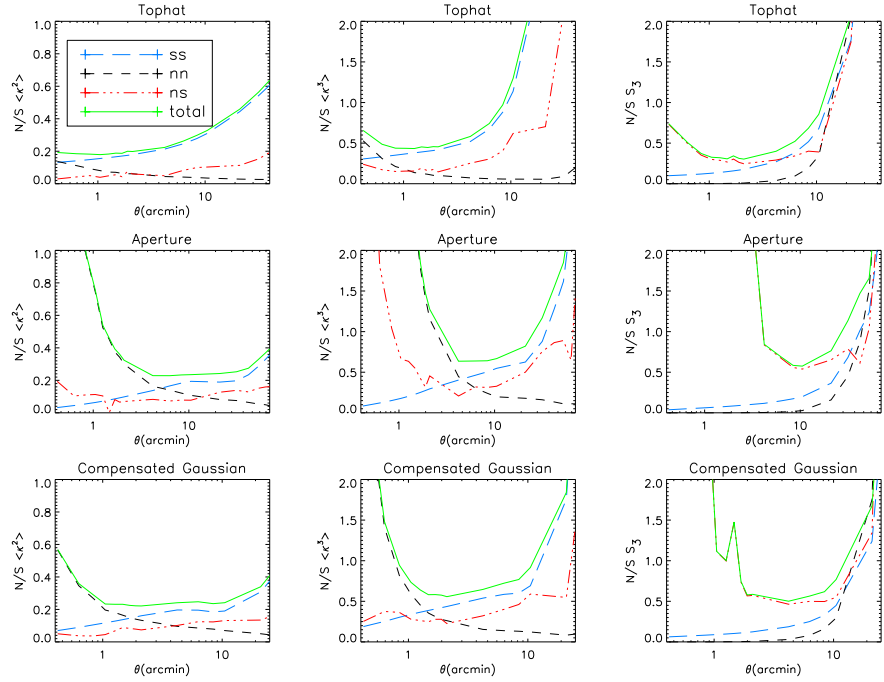}
\caption{The noise-to-signal ratio for the cosmic variance only in
blue (long dashed) line, statistical noise only in black (short
dashed) line, the mixed term in red (dash-dotted) line and the total
noise in green (solid). The $\langle \kappa^2 \rangle$, $\langle
\kappa^3 \rangle$ and $S_3$ measurements were calculated for a
simulated 12.84 deg$^2$ data smoothed with top-hat, aperture and
Compensated Gaussian (from top to bottom).} \label{fig:noisecontribution}
\end{figure}

Figure  \ref{fig:noisecontribution} shows the relative contribution
of different terms in the covariance matrix. The noise-to-signal
ratio for the individual components of the covariance matrix are
shown. The blue (long dashed) line is the signal-signal which is the
result of the cosmic variance only. The noise-noise term is shown
with black (short dashed) line. The mixed term was derived from ${\bf
C}-{\bf C}_{\rm ss}-{\bf C}_{\rm nn}$ and is shown in red
(dash-dotted) line and the green (solid) line shows the total noise
over signal ratio. As expected the finding was that small scales were
dominated by statistical noise and the large scales by cosmic
variance, where the signal is low. Interestingly, the mixed noise
term is relatively negligible for the two- and three-point
statistics, although it is clear that future high precision surveys
will have to take it into account. Surprisingly, the mixed noise
term is strongly dominant for the skewness.

In agreement with \cite{Zhang03} a range of
optimal angular scales (between one arcminute and half a degree) was found for
which the total noise affecting the two-point shear statistics is minimal, this is
also the case for the three-point statistics.

\subsection{\label{sec:fixedtime}Wide and Shallow versus Deep and Narrow}
Many of the future lensing surveys will have a limited observing
time and a full sky coverage will not be possible. The question will
arise whether a deep and narrow survey performs better than a large
and shallow survey. Therefore, it is important to quantify what is the
optimal balance between survey size and depth, given a fixed
observing time. It is expected that very shallow surveys would
provide a poor weak lensing measurement due to the small lensing
efficiency for nearby sources, and deep-narrow surveys will be
limited by cosmic variance. The trade-off between those radically
different survey  designs  must include a proper estimate of the
amplitude of the lensing signal and shot noise as function of survey
depth.

The relation between limiting magnitude and survey area for a fixed
observing time was derived from the algorithm developed in
\cite{Bernstein01}. The galaxy number density was obtained by
selecting galaxies which signal-to-noise detection level was larger
than 7 and which are also well resolved for weak lensing studies
following the criteria given in Section \ref{sec:galaxynumber}. Table
\ref{tab:sizelimmag} shows the survey area and limiting magnitude
for each case investigated here.

The likelihood is given by:

\begin{equation}
\mathcal{L}=\exp\left[{-\frac{1}{2}(d-m)^T*{\bf C}^{-1}*(d-m)}\right]
\label{eq:L}
\end{equation}
where $d$ is the data and $m$ is the theoretical model. {\bf C}$^{-1}$ is
the inverse covariance matrix over all lines of sight. As described
in the previous section the covariance matrix was computed directly
by using the simulations and its inverse had been re-calibrated using
equation (\ref{alphadef}). The likelihood contours were performed in the
$\Omega_m$-$\sigma_8$ parameter space. $\Omega_m$ was varied between
0.1 and 1.0  with 0.05 intervals and $\sigma_8$ values were between
0.50 and 1.50 with 0.05 intervals.
\begin{table}
\centering
\begin{tabular}{l l l l l l l l l}
\hline
Area(deg$^2$)&1400&1150&900&514&257&115&45&20\\
\hline
$\rm m_{lim}$&22.5&23.0&23.5&24.0&24.5&25.0&25.5&26.0\\
\hline
$\rm n_{g}/arcmin^2$&2&5&9&14&22&28&37&45\\
\hline
$\rm GF$&1.8&4.0&5.0&2.5&2.4&1.5&1.3&--$^{\star}$\\
\hline
\end{tabular}
\caption{The area and i-band limiting magnitude and the corresponding
galaxy number density of different surveys with the same observing
time. The gain factor $\rm GF$ is the ratio between the $\Omega_m$ 1$\sigma$ width of the two-point statistics contours over that of the two- and three-point statistics joint contour. ($^\star$) Due to the truncated likelihood 1$\sigma$ contours the GF is not calculated for the deepest survey.} \label{tab:sizelimmag}
\end{table}

\begin{figure}
\begin{center}
\includegraphics[scale=0.8]{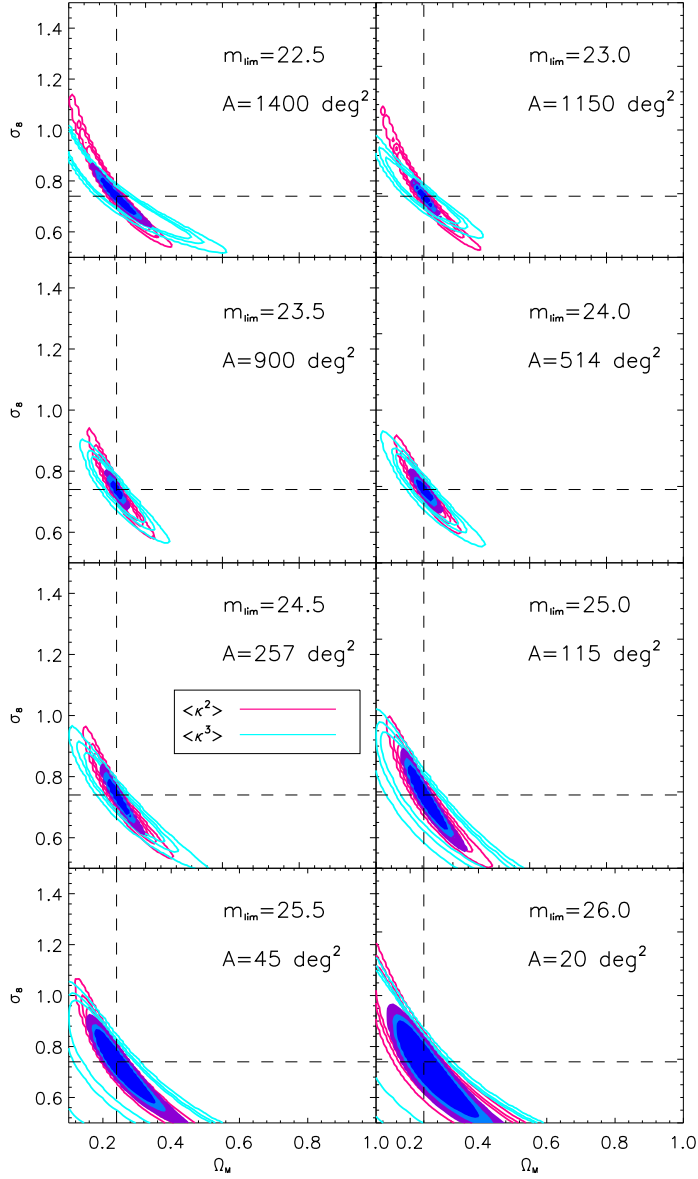}
\end{center}
\caption{ The likelihood analysis for various survey depths and
areas with fixed observing time for $\langle \kappa^2 \rangle$,
$\langle \kappa^3 \rangle $ smoothed with the top-hat filter. The
observing time is equal for all cases, while the survey area and
depth vary. Table \ref{tab:sizelimmag} shows the values for $\rm
m_{lim}$ with the corresponding survey areas. The pink (dark grey)
contours indicates the 1$\sigma$, 2$\sigma$ and 3$\sigma$ errors for
the $\langle \kappa^2 \rangle$ statistics and the cyan (light grey)
contours are the same for the $\langle \kappa^3 \rangle$. The
covariance matrix contains both the cosmic variance and the
statistical noise. Here the joint likelihood shown in filled
contours is calculated  by taking into the account the  $\langle
\kappa^2 \rangle$-$\langle \kappa^3 \rangle$ correlations at
different scales.} \label{fig:fixed}
\end{figure}

Figure \ref{fig:fixed} shows the pink (dark
gray) contours for $\langle\kappa^2\rangle$ and cyan (light grey)
for $\langle\kappa^3\rangle$ likelihood for top-hat filter. The filled contours show
the joint $\langle\kappa^2\rangle$-$\langle\kappa^3\rangle$
likelihood. The $\langle\kappa^2\rangle$ and
$\langle\kappa^3\rangle$ contours become more degenerate for deep
and narrow surveys, whereas  for wide and shallower surveys it appears clearly that
the $\langle\kappa^2\rangle$ and $\langle\kappa^3\rangle$ likelihood contours have a
different orientation in the  $\Omega_m$-$\sigma_8$ plane, which explains
why the joint analysis works better for wide and shallow surveys.
One can see indeed that the individual two- and three-point statistics contours
for the wide and shallow surveys become large again due to a larger
noise, but the joint analysis remains competitive. This could be
attributed to the larger sensitivity of the three-point statistics to
non-linear effects for shallow surveys as a consequence of the
projection of mass along the line-of-sight (i.e. identical angular
scale probes more non-linear scales for shallow rather than deep
survey). For the joint two- and three-point statistics analysis, the
medium depth surveys ($\rm m_{lim}=$23.5 or 24.0) appear optimal. It
is clear that for a fixed observing time, our results favor the
medium shallow-wide surveys.
The gain factor $\rm GF$ is defined as the ratio of the 1$\sigma$ error width of the $\langle\kappa^2\rangle$ contours over that of joint $\langle\kappa^2\rangle$-$\langle\kappa^3\rangle$ measurements which quantifies the improvement when the joint statistics is considered.
The values of the GF corresponding to the likelihood contours of figure \ref{fig:fixed} are shown in table \ref{tab:sizelimmag}.

\begin{figure}
\begin{center}
\includegraphics[scale=0.85]{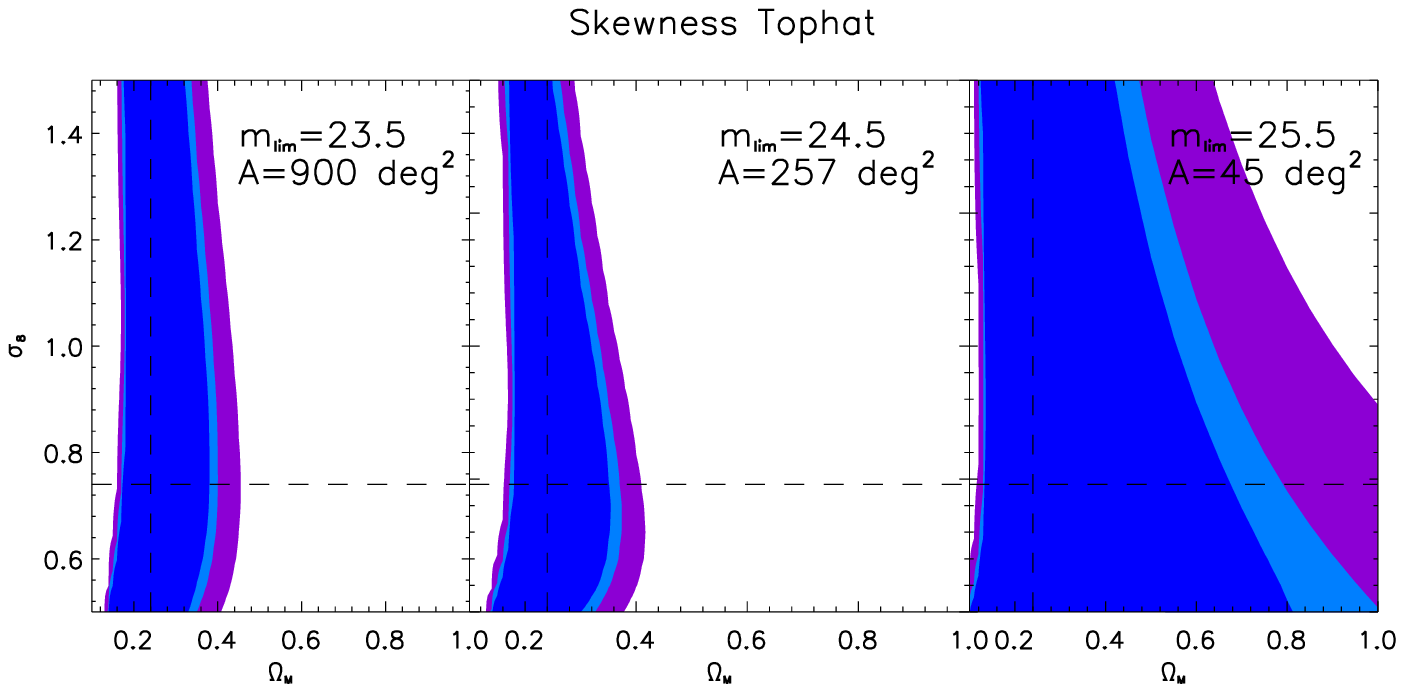}
\end{center}
\caption{ The likelihood analysis for various survey depths and
areas with fixed observing time for skewness $S_3$ smoothed with the
top-hat filter. The observing time is equal for all cases, while the
survey area and depth vary. Table \ref{tab:sizelimmag} shows the
values for $\rm m_{lim}$ with the corresponding survey areas. The
covariance matrix contains both the cosmic variance and the
statistical noise. The skewness measurements are most optimal for
shallower surveys.} \label{fig:skewlimmag}
\end{figure}

Unfortunately, the skewness of the convergence, defined in
equation (\ref{eq:s3}), does not appear to yield as powerful constraints as the combined two-
and three-point statistics.
 Figure \ref{fig:skewlimmag} shows the error
contours using $S_3$ for three choices of limiting magnitude and
survey area. The observing time here was fixed, like for
the previous analysis. As expected, the dependence on $\sigma_8$ is
very weak, but one can see that the width of the contours
along the $\Omega_m$ axis is much larger than the $\Omega_m$
constraints one gets from the joint analysis shown in Figure
\ref{fig:fixed}. Following the same trend as joint $\langle \kappa^2
\rangle $-$\langle \kappa^3 \rangle $ likelihood results shown in figure
\ref{fig:fixed}, the medium depth surveys lead to the most optimal
skewness measurement.
The  constraints for the shallower surveys (i.e.
$\rm m_{lim}=$22.5 and 23.0) are not shown here. Those surveys give poor cosmological
constraints, as the mixed $\bf{C_{ns}}$ term of the covariance  at
the scales of interest becomes large. Overall, the skewness does not appear to be as attractive a statistic to break the $\sigma_8$-$\Omega_m$ degeneracy as previously advocated (\cite{Bernardeau97} and \cite{waerbeke99}).
Measuring the skewness
on the current and near future lensing surveys will be very
challenging, and it is clear that a large fraction of the sky is
needed in order to bring the noise contributions to a low enough level
for precision cosmology.

\begin{figure}
\includegraphics[scale=0.6]{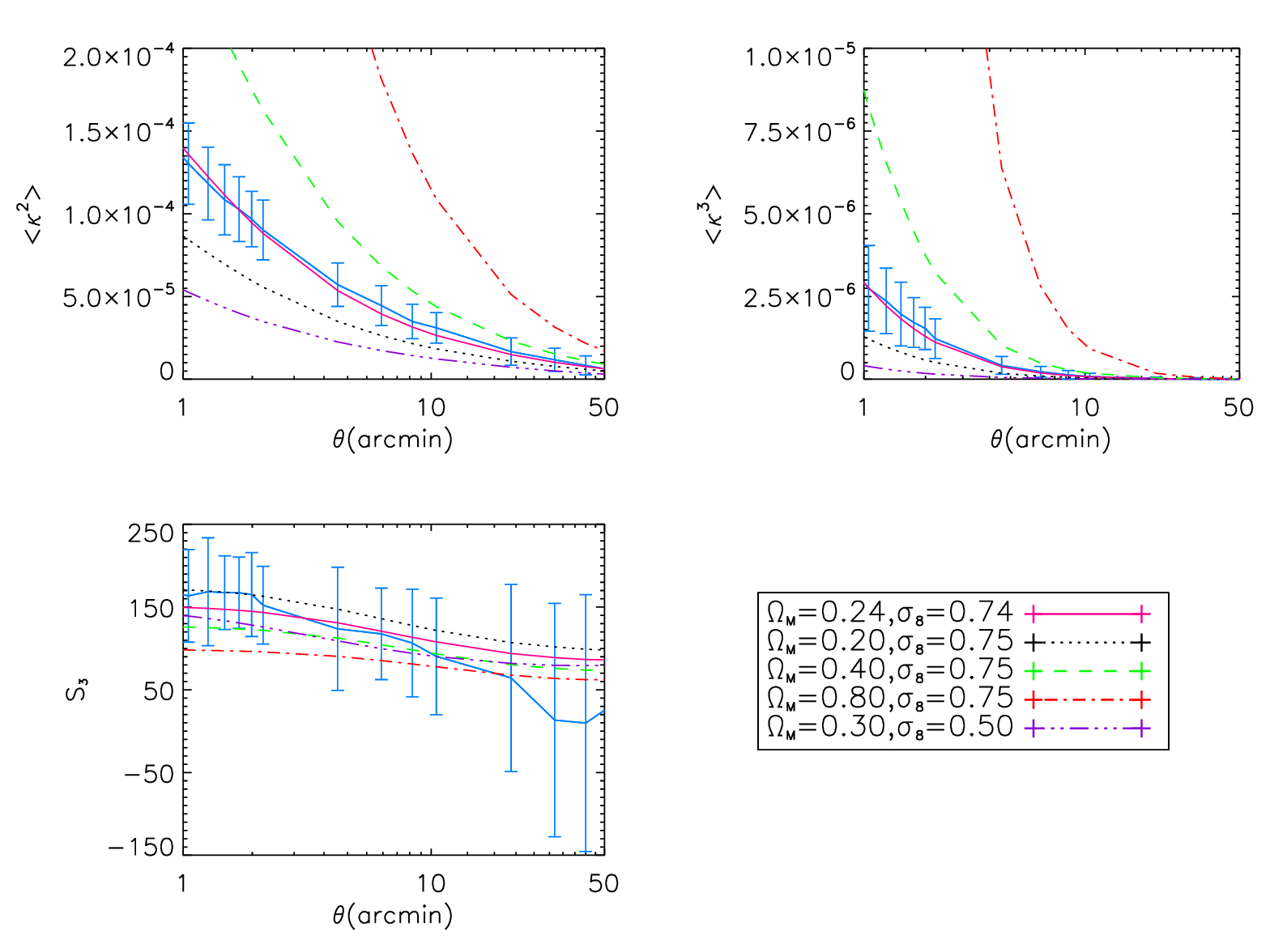}
\caption{The comparison between the measured values of $\langle
\kappa^2 \rangle $, $\langle \kappa^3 \rangle $ and $S_3$ and
different cosmological models over a survey area of 12.84
deg$^2$.The blue line shows the measured data points, and the pink
(solid) line is the fiducial model. The black (dotted),  green
(dashed) and red (dash-dotted) lines are models with the same
$\sigma_8=0.75$ and values of $\Omega_m=0.20,0.40$ and $0.80$
respectively, while the purple (dash-dot-dotted) line corresponds to
a model with  $\Omega_m=0.30$ but  $\sigma_8=0.50$. The plots show
that the measurement of $\langle \kappa^2 \rangle $ and $\langle
\kappa^3 \rangle $ are much more sensitive to the $\Omega_m$,
$\sigma_8$ parameters than the  skewness $S_3$. This is why we can
not currently constrain the $\Omega_m$-$\sigma_8$ plane with
skewness measurements.} \label{fig:s3withmodels}
\end{figure}

The reason why the skewness is hard to measure lies in the fact that
the variation of the skewness amplitude for different $\Omega_m$
models is largely absorbed by the cosmic variance of this estimator.
This is not the case for the two- and three-point statistics  taken separately.
Figure \ref{fig:s3withmodels}  shows the comparison between various predicted cosmological models and the measurements from the simulations.
$\langle \kappa^2
\rangle $, $\langle \kappa^3 \rangle $ and $S_3$ were measured for survey area of
12.84 deg$^2$ of limiting magnitude of 24.5 over the 60 lines of
sight. The blue line shows the measured data points; the errorbars
contain both cosmic variance and statistical noise. The pink (solid)
line is the fiducial model ($\Omega_m$=0.24, $\Omega_\Lambda$=0.76
and $\sigma_8$=0.74). The black (dotted), green (dashed) and red
(dash-dotted) lines are models with the same $\sigma_8=0.75$ and
values of $\Omega_m=0.20,0.40$ and $0.80$ respectively, while the
purple (dash-dot-dotted) line corresponds to a model with
$\Omega_m=0.30$ but  $\sigma_8=0.50$. The plot  shows that the
measurement of $\langle \kappa^2 \rangle $ and $\langle \kappa^3
\rangle $ are much more sensitive to the $\Omega_m$, $\sigma_8$
parameters than the skewness $S_3$, therefore their ability to
separate various cosmological models is stronger.

\section{\label{sec:CFHTLS}Canada-France-Hawaii Legacy Survey Three-point Statistics Predictions}

The Canada-France-Hawaii Telescope Legacy Survey covers
170 deg$^2$ in four patches \cite{Fu08}. Measurements of the
two-point cosmic shear statistics have been published using the
first year (\cite{Hoekstra06}, \cite{Semboloni06} and
\cite{Benjamin07}) and third year data release \cite{Fu08} in
addition to studies of galactic scale dark matter halos
\cite{Parker07}. In this study the expected improvement
for cosmological parameter constraints, using a combination of
two- and three-point lensing statistics on the completed
CFHTLS-wide survey was determined.  A mock CFHTLS-wide survey type of $170$ deg$^2$ was generated using a limiting magnitude of $\rm {m}_{\rm lim}=24.5$ (i-band) with $n_g$=22 galaxies per arcmin$^2$,  $z_{\rm med}$=0.91 and
$\sigma_\epsilon$=0.44 and the potential contamination by residual systematics was ignored.

\begin{figure}
\begin{center}
\includegraphics[scale=0.9]{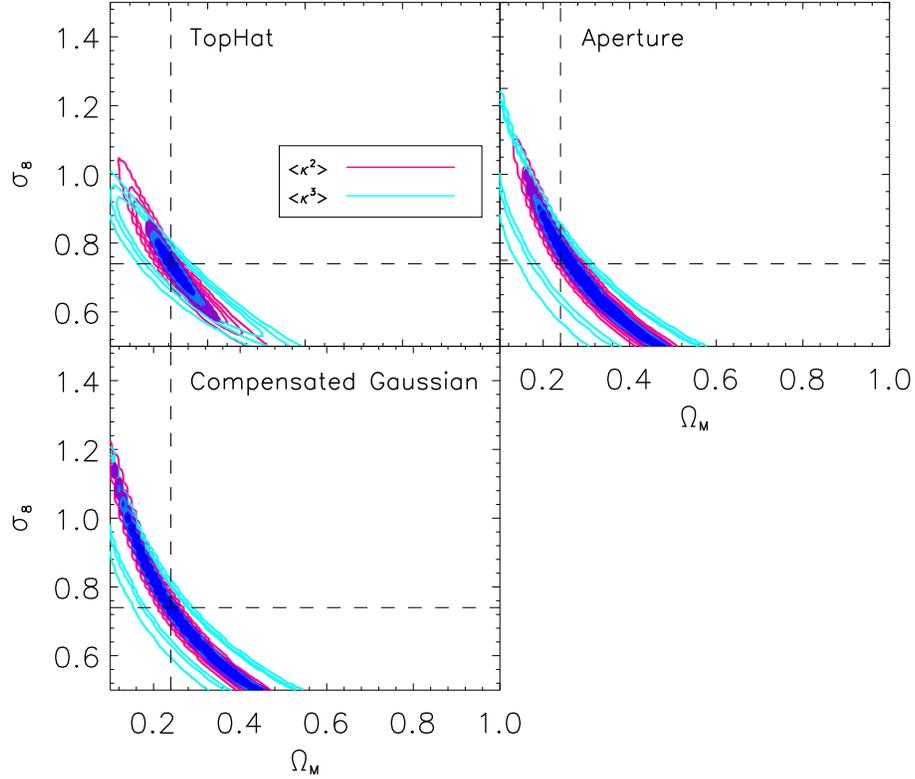}
\end{center}
\caption{The cosmological constraints on $\Omega_m$-$\sigma_8$ plane
obtained with different smoothing filters.  The contours show the
$\langle \kappa^2 \rangle $ and $\langle \kappa^3 \rangle$ joint
likelihood forecast based on CFHTLS completed area. The survey area
is 170 deg$^2$ and the limiting magnitude is 24.5 with the full
redshift distribution. Here the pink (dark grey) contours show the
$\langle \kappa^2 \rangle $ and the cyan (light grey) contours show
the $\langle \kappa^3 \rangle $ constraints. The filled contours
correspond to the 1$\sigma$, 2$\sigma$ and 3$\sigma$ errors for the
joint likelihood. The fiducial model used is a $\Lambda$CDM with
$\Omega_m=0.24$ and $\sigma_8=0.74$. The  degeneracy direction of the $\langle \kappa^2 \rangle
$ and $\langle \kappa^3 \rangle$ likelihood is different (especially when the maps are smoothed with
Compensated Gaussian filter) so their joint likelihood results in a
tighter constraints on the parameters. The joint likelihood here is
calculated by taking into account the cross correlations between
$\langle \kappa^2 \rangle $ and $\langle \kappa^3 \rangle$ at all
scales.} \label{fig:cfhtlsjointlike}
\end{figure}

Figure \ref{fig:cfhtlsjointlike} shows the $\Omega_m$ and $\sigma_8$
error contours from a joint measurement of $\langle \kappa^2 \rangle
$ and $\langle \kappa^3 \rangle$ for the three filters used in this
study.
 It is clear that the top-hat
filter leads to a more significant degeneracy breaking between
$\Omega_m$ and $\sigma_8$, which can be understood by the fact that
this filter preserves modes with wavelengths larger than the
smoothing size, while the aperture filters are not sensitive to
large scale variations.
The joint two- and three-point analysis of the completed CFHTLS-Wide will constrain  $\Omega_m$ and $\sigma_8$  to 17\% and 10\% respectively.  This corresponds to a gain factor (GF) of $\sim$ 2.5 (for $\Omega_m$) and $\sim$ 2.1 (for $\sigma_8$) improvement on the two-point analysis alone when the top-hat filter is used.
It is interesting to compare Figure
\ref{fig:cfhtlsjointlike} to a generalized $\chi^2$ approach which
can serve to quantify the performance of the different filters. The
generalized $\chi^2$ is defined as:

\begin{equation}
\frac{S}{N}=\sqrt{d^T * {\bf C}^{-1} * d}
\end{equation}
where ${\bf C}$  is the covariance matrix of the statistics under
consideration. The results for a $12.84$ deg$^2$ survey and
limiting magnitude $\rm m_{lim}=24.5$ are shown in table
\ref{tab:chi2}, and they indicate that for two-point statistics the
different filters are equivalent. The top-hat filter outperforms the
aperture filters for the three-point statistics. It is a direct
illustration that top-hat preserves small and large scale modes, and
it is therefore more sensitive to non-linear effects. This
invalidates the fact that the Compensated Gaussian filter is the most efficient measure of
the skewness of the convergence \cite{Zhang03}. The reason lies in the fact that for a fair comparison the maximum smoothing scale for Compensated Gaussian filter may not exceed a third of top-hat and aperture smoothing radii. This can be seen by looking at the equations which define the shape of the filters (see Section \ref{sec:filters}).
\begin{table}[!h]
\centering
\begin{tabular}{l l l l }
\hline
S/N&$\langle \kappa^2 \rangle $&$\langle \kappa^3 \rangle $&$S_3$\\
\hline
Top-hat& 6.19&2.68&5.45 \\
Aperture &6.05&1.61&2.17\\
Compensated Gaussian & 6.93&1.88&3.24 \\
\hline
\end{tabular}
\caption{The generalized $\chi^2$ results for top-hat, aperture and
Compensated Gaussian filters. The full covariance matrix is that of
the 12.84 deg$^2$ maps. The data $d$ is from the $\kappa$-maps
smoothed with top-hat, aperture and Compensated Gaussian filters. The
correlation between the scales are contained in the signal-to-noise
ratio.} \label{tab:chi2}
\end{table}

\begin{figure}
\begin{center}
\includegraphics[scale=0.9]{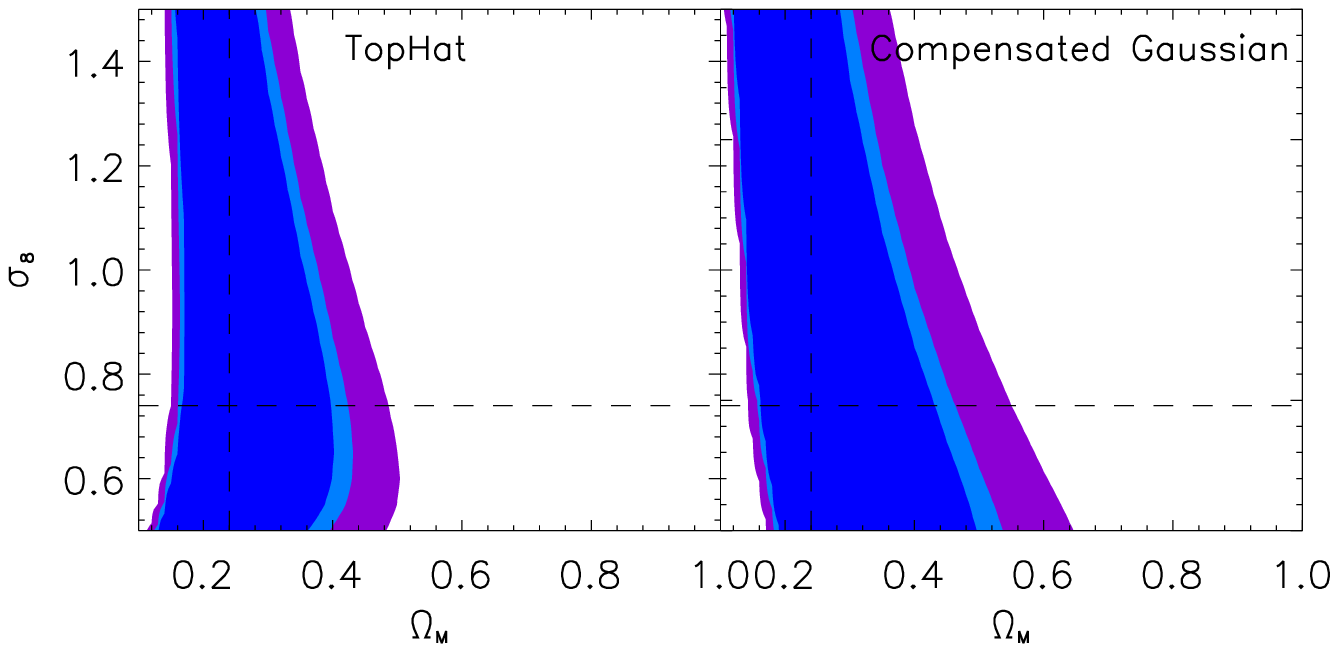}
\end{center}
\caption{The skewness $S_3$ likelihood analysis for the CFHTLS-wide
predictions. The smoothing filters top-hat and Compensated Gaussian
are used. The aperture filter does not provide any constraint on the
$\Omega_m$-$\sigma_8$ plane for the given survey characteristics.
The covariance matrix contains both the cosmic variance and the
statistical noise.} \label{fig:cfhtlsS3}
\end{figure}

The joint $\langle \kappa^2 \rangle $-$ \langle \kappa^3 \rangle$
likelihood analysis with top-hat, aperture and Compensated Gaussian
filters proved to be promising, whereas the skewness which is in principle a
very interesting statistic inferred very weak cosmological constraints even for the current largest weak lensing survey at 170 deg$^2$.  Figure \ref{fig:cfhtlsS3} shows
skewness likelihood contours obtained using both top-hat and Compensated
Gaussian filter for CFHTLS-like survey confirming what stated above about the
poor efficiency of the skewness.

\begin{figure}
\centering
\subfigure[]{
\includegraphics[scale=0.8]{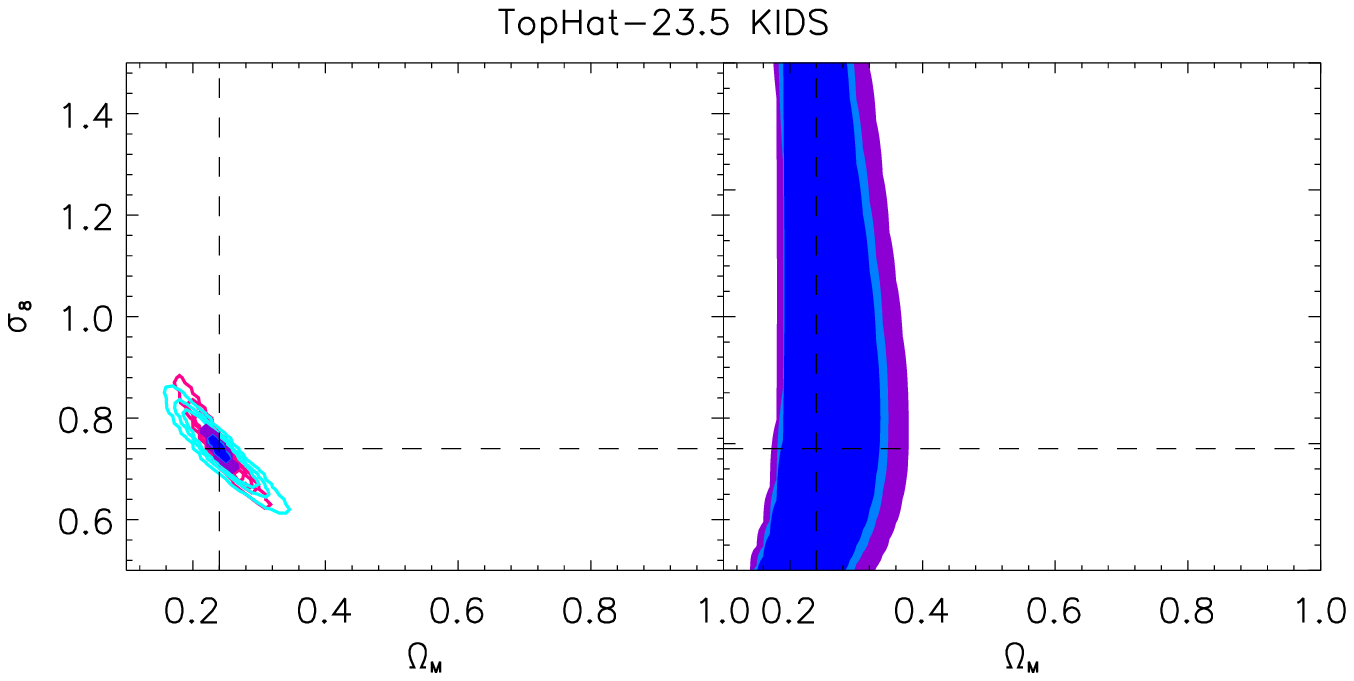}
\label{fig:kids23.5}
}
\subfigure[]{
\includegraphics[scale=0.8]{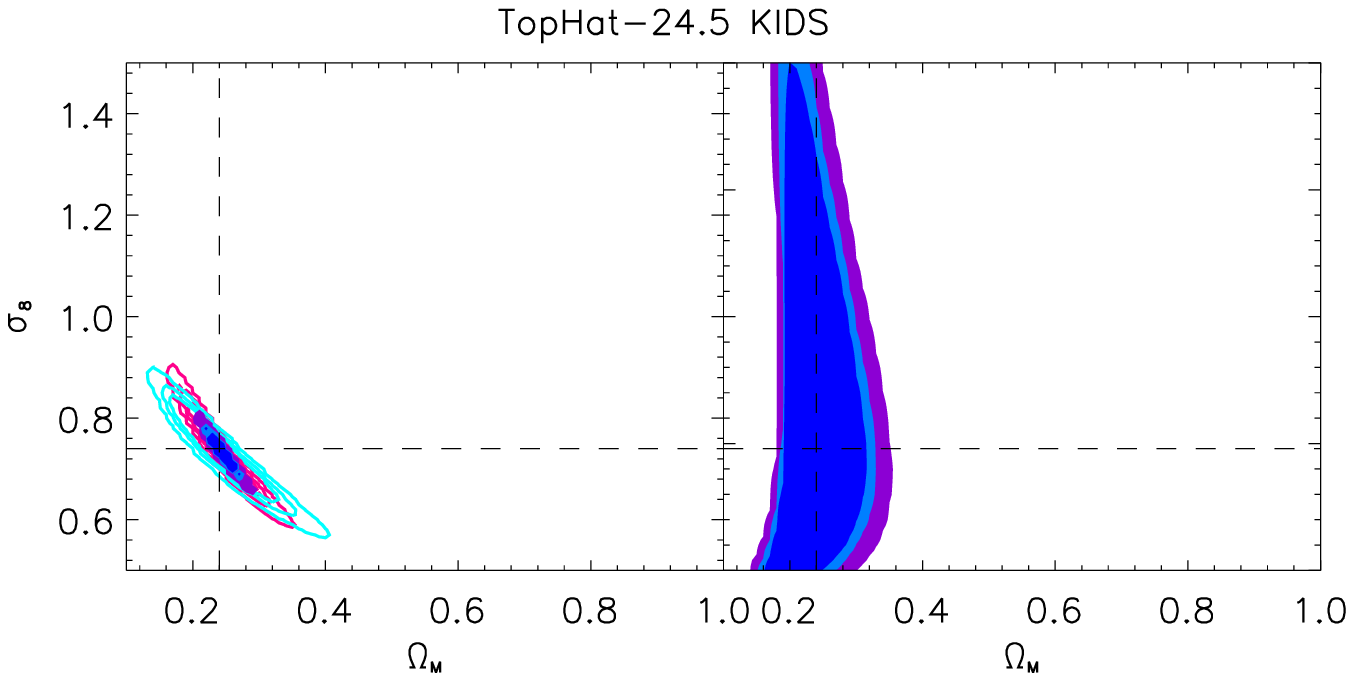}
\label{fig:kids24.5}
}
\caption[]{The comparison between a KIDS-like survey at two limiting magnitudes.
The right panels show the likelihood contours of  $\langle \kappa^2 \rangle$ and $\langle \kappa^3 \rangle $ smoothed with the top-hat filter. The left panels are the  skewness contours. The survey area for the  panel  \subref{fig:kids23.5} is 1500 deg$^2$ as is planned for the KIDS survey with $\rm m_{lim}$=23.5. In panel \subref{fig:kids24.5} the observing time is kept the same  and the survey area if adjusted to 450 deg$^2$ for $\rm m_{lim}$=24.5.}
\end{figure}

One of the forthcoming weak lensing surveys is the KIlo Degree Survey
(KIDS area of 1500 deg$^2$ at $\rm m_{lim}$=23.5). We performed for the KIDS survey the same analysis as for the CFHTLS-Wide to forecast
the accuracy of the likelihood constraints using two- and three-point
shear statistics. Moreover, for comparison the calculations were repeated for a survey with the
same observing time needed for the KIDS survey but different total
area and depth.  The results establish which survey design  would be the most
optimal to infer constraints using the joint two- and and three-
point shear statistics. The expected likelihood contours  for the
complete KIDS survey are shown in the panel \ref{fig:kids23.5},
whereas panel \ref{fig:kids24.5} shows the same results when a deeper ($\rm m_{lim}$=24.5) and narrower (area=450
deg$^2$) survey given the same observing time was considered. As expected from
Figure \ref{fig:fixed} the shallower KIDS gives better results for
the joint likelihood, but the skewness would be slightly better
measured from the deeper ($\rm m_{lim}$=24.5) survey.

\section{\label{sec:conclusion}Conclusion and Discussion}

We studied how useful the measurement of the two- and three-point shear statistics can be to derive cosmological constraints under
realistic observing conditions. One of the limitations of the
previous work on this topic was the disconnection between the source
redshift distribution and the survey depth under consideration.
Here, a set of ray-tracing simulations was populated with source
galaxies that follow a redshift distribution and galaxy number
density calibrated from real data.

We then investigated how well the parameters $\sigma_8$ and
$\Omega_m$ can be measured with different smoothing filters for
different survey depths. For a fixed observing time, the results of the study favored
the medium depth and width survey over shallower and wider or deeper
and narrower surveys. There is an optimal survey depth versus size
for which the source density (survey depth) and cosmic variance
(survey area) are balanced, which turns out to be very close to the
CFHTLS depth. Our results can be applied also to surveys covering a
large fraction of the sky with no limitation on the observing time
(e.g. PanSTARRS and LSST) by a simply rescaling the covariance
matrices. They can also be extended to space data if the amplitude
of ellipticity noise and galaxy number counts are adjusted according
to space observations (this is particularly relevant for a wide
field space imager like the one by JDEM).

We find that the lensing statistics are best measured at scales
between 1 to 30 arcminutes, where the contribution of statistical
noise, cosmic variance and the mixed term are minimal. We also find
that the different smoothing filters give similar results although
the top-hat appears to include more modes and is therefore slightly better than the others. Combining  $\langle \kappa^2
\rangle $ and $\langle \kappa^3 \rangle$ is promising to achieve cosmological constraints in the $\Omega_m$-$\sigma_8$
parameter space. On the contrary the skewness of the
convergence does not appear able of breaking the
degeneracy between $\sigma_8$ and $\Omega_m$ as initially expected (\cite{Bernardeau97} and \cite{waerbeke99}). The reason is that the cosmic variance on $S_3$ is
comparable to the difference in lensing signal amplitude for the
different cosmological models of interest. Only very large surveys
will be able to measure the skewness accurately.

We forecasted the cosmological constraints for the CFHTLS-wide
completed survey finding that the combination of two- and
three-point functions on the CFHTLS will greatly enhance the
measurement of $\sigma_8$ and $\Omega_m$. A similar calculation
showed the potential precision achievable with the future KIDS
survey.

This study has some limitations which will be investigated in future
work. One of them is the fact that the source galaxies are clustered
in three dimensional space which overlap with lens redshift
distribution (problem known as the source clustering problem,
\cite{Bernardeau98}). This effect leads to a change in the skewness
of the convergence (by as much as $25\%$), and its impact on the
three-point statistics has not been evaluated yet. A recent study
also showed potential impact on the two-point statistics, although
at a moderate level (\cite{Forero07}). Another limitation is the
potential impact of intrinsic alignment on three-point statistics.
This is particularly relevant for shallow surveys such as PanSTARRS
or KIDS (\cite{Semboloni08}). This effect should be taken into the
account as well. We would be able to investigate these two
complications with ray-tracing simulations which include galaxies in
dark matter halos; this can be realized by the use of semi
analytical models such as the ones described in \cite{Forero07}.

\section{Acknowledgement}
We used the CFHTLS photometric redshift catalogues generated by
Olivier Ilbert. All N-body simulations were performed on the Canada
Foundation for Innovation funded CITA Sunnyvale cluster. The
analysis of the simulations was performed on WestGrid computing
resources, which are funded in part by the Canada Foundation for
Innovation, Alberta Innovation and Science, BC Advanced Education,
and the participating research institutions. LVW acknowledges
support from NSERC, CFI and CIAR. ES acknowledges the support of the
Alexander von Humboldt foundation. We are thankful to Gary Bernstein
for the ETC (Exposure Time Calculator) software. We are also
thankful to Martin Kilbinger and Alan Heavens for helpful
discussions. 


\end{document}